\input{aipcheck}

\documentclass[
    ,final            
  ]
  {aipproc}

\layoutstyle{8x11single}
\usepackage{wasysym}

\begin{document}

\title{Overview of the European Underground Facilities}

\classification{01.52.+r, 07.85.Nc, 96.50.S-, 95.35.+d}
\keywords      {Deep underground science laboratories, Europe, astroparticle physics, 
$\gamma$-ray spectrometry}

\author{L.~Pandola}{
  address={INFN, Laboratori Nazionali del Gran Sasso, SS 17~bis km 18+910, I-67100 
Assergi (AQ), Italy}
}

\begin{abstract}
Deep underground laboratories are the only places where the extremely low background 
radiation level required for most experiments looking for rare events in physics 
and astroparticle physics can be achieved. Underground sites are also the most 
suitable location for very low background $\gamma$-ray spectrometers, able to assay 
trace radioactive contaminants. Many operational infrastructures are 
already available worldwide for science, differing for depth, dimension and rock 
characteristics. Other underground sites are emerging as potential new laboratories. \\
In this paper the European underground sites are reviewed, giving a particular emphasis 
on their relative strength and complementarity. A coordination and integration effort 
among the European Union underground infrastructures was initiated by the EU-funded 
ILIAS project and proved to be very effective.
\end{abstract}

\maketitle


\section{Introduction}
Underground laboratories provide the low radioactive background environment which is necessary for 
most key experiments in the field of rare events and astroparticle physics. 
While some experiments (e.g. dark matter search, neutrinoless double beta decay, solar 
neutrinos) require very deep sites ($> 2$~km w.e.), other experiments - neutrino 
beam, proton decay - can be accommodated in shallower sites, since their 
requirements on the residual muon flux are weaker. \\

The low-radioactivity environment of underground laboratories is extremely beneficial 
also for ultra low-background $\gamma$-ray spectroscopy. In fact experiments must be 
built out of very radiopure materials, and  most of them place very stringent 
requirements on radioactive contaminants. Therefore, materials and components to be 
used for experiments must be screened by ultra-sensitive Ge spectrometers in order to 
be qualified for actual use. The background level required for HPGe spectrometers to 
be sensitive to radioactive contamination down to a few mBq/kg can be 
practically achieved in underground laboratories only. In principle, ultra-sensitive $\gamma$-ray 
spectrometry does not require very deep sites since after a few hundreds of m w.e. 
the total background is dominated by the radioactivity of the passive shielding. 
Nevertheless, virtually all underground laboratories host one or more 
Ge spectrometers as a screening facility for ultra low-background radioactivity 
assays. Furthermore, ultra-sensitive $\gamma$-ray spectrometers in underground 
laboratories are used for a variety of additional scientific applications, e.g.  
environmental monitoring and radio-dating, as well as for the development of new 
detector technologies.\\

\begin{figure} \label{fig:map}
  \includegraphics[height=.6\textwidth]{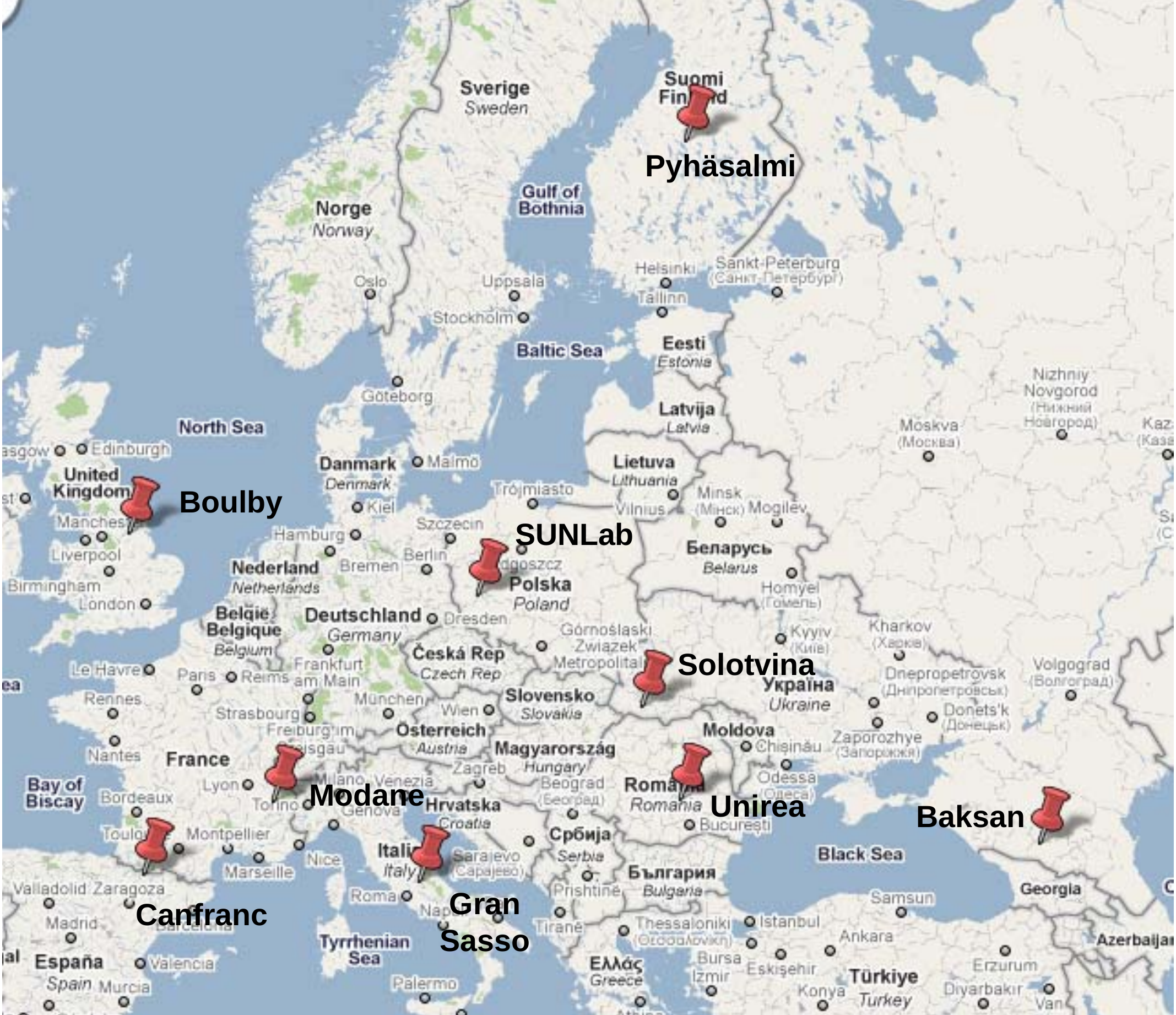} 
  \caption{Position of the European underground sites described in this paper.}
\end{figure}

Several underground facilities are presently operational in Europe, differing for 
depth, dimension and scope. At the moment there are nine cavities with overburden 
larger than about 600 m w.e. that can be used for science. They are  
displayed in Fig.~\ref{fig:map} and their main facts are 
summarized in Table~\ref{tab:maintable}. The sites will be described in more detail in 
the following Sections, based on Refs.~\cite{bet07,coc09}. 
Nevertheless only a fraction of these sites can be regarded as an operational 
``science infrastructure'' (with technical and technological support, 
safe access, supplies, etc.) and is able to host full-scale 
astroparticle physics experiments. Other sites are either emerging, namely are cavities in the potential 
state to become a full infrastructure in the future, or are located at shallower depth and are 
mainly targeted to $\gamma$-ray spectrometry or to specific projects, as 
LAGUNA\footnote{LAGUNA~\cite{laguna} is a Design Study (DS) project, funded by the European Commission 
under the $7^{th}$ Framework Program focusing on the design of a new European 
underground infrastructure able to host large-scale experiments (tens of ktons up to 1 Mton) 
on low-energy neutrino astronomy and proton decay search.}. \\
There are five deep underground sites in Europe that can be regarded as operational 
scientific infrastructures: one in Russia (Baksan Neutrino Observatory) and four in 
the European Union: Gran Sasso National Laboratory (LNGS), Italy; Modane Underground Laboratory 
(LSM), France; Boulby Palmer Laboratory, United Kingdom; Canfranc Underground Laboratory (LSC), Spain. 
All these laboratories are established since more than ten years, they have hosted key 
experiments and have a rich scientific program for the future, encompassing dark matter searches,
neutrinoless double beta decay, neutrino physics and astroparticle physics. \\
The four laboratories in Western Europe took part in the ILIAS (Integrated Large Infrastructures 
for Astroparticle Science) project~\cite{ilias} between 2004 and 2009, which was funded 
by the European Commission under the 6$^{th}$ Framework Program for a total amount of 
7.5 million euro. The ILIAS project was beneficial for a closer coordination of the activities 
of the laboratories and for the optimization of the available resources, taking into account the 
different characteristics of the infrastructures. The common activity 
encompassed coordination for environmental background measurement and control, safety 
procedure and outreach. A program of Trans-National access was active within ILIAS to favor 
visits of new scientific groups in the underground infrastructures. The underground 
site at the Pyh\"asalmi mine (Finland), which is not an operational underground 
infrastructure yet, did not take part in the ILIAS Transnational Access program, but it was 
included in the coordination activity related to background measurements and safety 
procedures. \\
The long-term continuation of the coordination among the four main EU deep underground 
infrastructures was favored by ASPERA~\cite{aspera} after the end of the ILIAS 
program\footnote{ASPERA is a network of national government agencies, funded by the 
European Commission under the 7$^{th}$ Framework Program, which is responsible for 
coordinating national research efforts in Astroparticle Physics in Europe.}. A new 
collaboration agreement called EULab was signed in 2009 by the agencies that run 
the laboratories.\\
\begin{table}
\begin{tabular}{l|lllll}
\hline
  & \tablehead{1}{c}{b}{Surface \\ profile}
  & \tablehead{1}{c}{b}{Surface \\ (m$^{2}$)}
  & \tablehead{1}{c}{b}{Minimum depth \\ (m.w.e.)}
  & \tablehead{1}{c}{b}{Muon flux \\ (m$^{-2}$s$^{-1}$)}
  & \tablehead{1}{c}{b}{Neutron flux \\ (m$^{-2}$s$^{-1}$)} \\
\hline
Baksan (Ru) & Mountain & ca. 3000 & 850$\to$4800 & $3.03 \cdot 10^{-5}$ & 
$0.23 \cdot 10^{-2}$ ($1-11$~MeV) \\
\hline
Modane (F) & Mountain & ca. 400 & 4000 (min) & $5.76 \cdot 10^{-5}$ & 
$1.1 \cdot 10^{-2}$ ($> 1$~MeV) \\
& & & 4800 (average) & & $1.9 \cdot 10^{-2}$ (thermal) \\
\hline
Gran Sasso (I) & Mountain & 17~000 & 3100 (min) & $2.87 \cdot 10^{-4}$ & 
$0.86 \cdot 10^{-2}$ ($> 1$~keV) \\
 & & & 3800 (average) & & $2.93 \cdot 10^{-2}$ ($< 1$~keV) \\
\hline
Boulby (UK) & Flat & 1500 & 2850 (min) &  $4.1 \cdot 10^{-4}$ & 
$1.72 \cdot 10^{-2}$ ($> 0.5$~MeV) \\  
 & & & 3450 (average) & & \\
\hline
Canfranc (E) & Mountain & ca. 1000 & $\approx$ 2400 (max) &  $2-4 \cdot 10^{-3}$ &
$3.82 \cdot 10^{-2}$ ($> 1$~MeV) at old site\\ 
\hline
Pyh\"asalmi (Fin) & Flat & ca. 1000 & 200$\to$2400 & N/A & N/A \\
\hline
Sieroszowice (Pol) & Flat & many 1000's & $\approx$ 2200 & N/A & N/A \\
\hline
Solotvina (Ukr) & Flat & 1000 & $\approx$ 1000 & $1.7 \cdot 10^{-2}$ & 
$< 2.7 \cdot 10^{-2}$ (thermal) \\
\hline
Unirea (Rom) & Flat & 70~000 (tot) & $\approx$ 600 & N/A & N/A \\ 
\hline
\end{tabular}
\caption{Summary table of the European underground sites available for science that are 
described in this paper. References for the laboratory characteristics and fluxes 
are reported in the text. Sites are ordered in decreasing depth.}
\label{tab:maintable}
\end{table}
\section{The ILIAS laboratories} \label{sec:eulab}
In the following subsections, the four ILIAS deep underground laboratories and the 
Pyh\"asalmi site will be described in more detail. Most information is taken from 
Refs.~\cite{bet07,coc09} and from the reports of the ILIAS N2 coordination 
network (``Deep underground science laboratories'') and JRA1 joint research activity 
(``Low background techniques for deep underground science''), available at 
\url{http://www-ilias.cea.fr}. Facilities are reported in chronological order 
since their establishment. 
\subsubsection{Laboratoire Subterrain de Modane (LSM), France}
The Modane underground laboratory is located under the Frejus mountain, in France, near the
Italian border. The infrastructure is run jointly by CNRS/IN$^{2}$P$^{3}$ and 
by CEA/DSM. It is the deepest underground infrastructure available in Western Europe, its 
minimum rock overburden being 4000 m w.e. (average overburden: 4800 m w.e.). 
The access to the laboratory is horizontal, through the single-way Frejus 
roadway tunnel. Intervention of the tunnel control is actually requested to stop 
the traffic and allow the entrance/exit of vehicles to the laboratory. The excavation 
started in 1979 and the first experiment - a 900 ton iron tracking calorimeter to 
search for proton decay (``Frejus experiment'') - was commissioned in 1982. The underground 
site is composed by a Main Hall, $30\times10\times11$~m$^{3}$, a 70~m$^{2}$ Gamma Spectrometry Hall 
hosting 13 HPGe detectors, and two secondary halls (18 and 21~m$^2$), totaling about 
400 m$^2$. Surface facilities located in the city of Modane include offices and a 
warehouse. \\  
The muon flux has been measured to be $5.76 \cdot 10^{-5}$~$\mu/($m$^2 \cdot$s)~\cite{ber89}, 
namely five muons per m$^2$ per day. The flux of fast neutrons (above 1~MeV) which has been 
re-measured in the framework of the EDELWEISS experiment is  
$(1.1 \pm 0.1) \cdot 10^{-2}$~n/(m$^2 \cdot$s)~\cite{lem06}. The thermal neutron flux is about 
$1.9 \cdot 10^{-2}$~n/(m$^2 \cdot$s)~\cite{sav10}. The laboratory has been also characterized 
in term of primordial radionuclides contained in the rock(concrete): 0.84(1.9)~ppm 
for $^{238}$U and 2.45(1.4)~ppm for $^{232}$Th. The Potassium content of the rock is 
6.9$\permil$ (= 213~Bq/kg of $^{40}$K). 
The average $^{222}$Rn content in the underground site is about 15~Bq/m$^3$ when 
ventilation is on (1.5 volumes per hour). Recently, an ``anti-radon'' facility was 
built underground which is able to produce Rn-free air (about 15~mBq/m$^3$) at a rate of 
150~m$^3$/h. The Rn-free air is necessary to meet the background specifications for 
the NEMO3 neutrinoless double beta decay ($0 \nu 2 \beta$) experiment. \\
At the moment, the laboratory is completely filled by two main experiments, EDELWEISS-II 
(dark matter search) and NEMO3 ($0 \nu 2 \beta$), and by a low-radioactivity counting 
facility. Other experiments having a smaller footprint, TGV-II ($0 \nu 2 \beta$ decay) and 
SHIN (nuclear physics), are also located in the laboratory. \\
The Gamma Spectrometry Hall hosts 13 HPGe low-background detectors, belonging to six 
different institutions. They are routinely used for radio-assay and qualification of 
materials to be used for the experiments but also for other applications, including 
environmental measurement and applications (wine dating, salt origin, etc.).\\
An extension of the laboratory of 60~000~m$^3$ (ULISSE project) is planned in 
coincidence with the excavation of a new safety tunnel, which started in September 2009. 
Two new large Halls (Hall A, $24\times100$~m$^2$, and Hall B, $18\times50$~m$^2$) are 
expected to be operational in 2013.
\subsubsection{Laboratori Nazionali del Gran Sasso (LNGS), Italy}
The Gran Sasso underground laboratory is one of the four national laboratories 
run by the INFN (Istituto Nazionale di Fisica Nucleare). It is located under 
the Gran Sasso mountain, in central Italy, and it is by far the largest 
underground laboratory in the world, serving the largest and most international 
scientific community. The vehicle access to the laboratory (cars and trucks) is through the 
A24 freeway (two-way) tunnel and does not require any specific safety 
procedure. \\
The total underground area is 17~000~m$^2$ (volume: 180~000~m$^3$). The 
laboratory encompasses three main halls (named A, B and C), each about 
$100\times20\times18$~m$^{3}$, plus ancillary tunnels, providing space for 
services, plants and smaller-scale experiments. After the initial proposal by the 
INFN President A.~Zichichi in 1979, the actual excavation started in 1982, in 
coincidence with the freeway 10-km tunnel, and was completed in 1987. The rock 
overburden of the laboratory is about 3100 m w.e. (minimum depth), or 
3800 m w.e. (average depth). Surface facilities are located in the town of 
Assergi and include offices, mechanical and electronics workshops, storage 
facilities, chemistry lab, computer and networking, canteen, conference rooms. The 
total personnel (physicists, engineers, technicians, administration) includes a 
permanent staff of about 80 and about 20 non-permanent positions.  Major civil engineering works 
have been accomplished between 2004 and 2007 to upgrade the safety conditions as well as  
the interface with the freeway system and the drinking water collection system.\\
The muon flux in the underground laboratory is 
$(2.87 \pm 0.03) \cdot 10^{-4}$~$\mu/($m$^2 \cdot$s)~\cite{agl98,ahl90}, corresponding to about 
one muon per 
m$^{2}$ per hour. Measured thermal and fast neutron fluxes are 
$2.93 \cdot 10^{-2}$~n/(m$^2 \cdot$s) below 1~keV and 
$0.86 \cdot 10^{-2}$~n/(m$^2 \cdot$s) above 1~keV, respectively~\cite{bel89}. 
The $^{222}$Rn specific 
activity in the underground halls is between 50 and 120 Bq/m$^3$ when the 
ventilation is on (one volume in 3.5 hours). The abundance of primordial 
radionuclides in the rock(concrete) is about 0.42(1.05)~ppm for $^{238}$U and 
0.062(0.656)~ppm for $^{232}$Th, respectively. The Potassium concentration in the 
Gran Sasso rock (Hall B) amounts to 160~ppm, corresponding to 4.9~Bq/kg of $^{40}$K; the 
Hall A rock has a much higher Potassium concentration, approx. 7$\permil$.\\ 
The current experimental program at LNGS is very rich, including CERN to Gran Sasso 
beam experiments (OPERA and ICARUS); neutrinoless double beta decay searches 
(CUORE, GERDA and COBRA); dark matter searches (DAMA/Libra, WArP, CRESST and XENON); 
solar and geo-neutrinos (Borexino); supernova neutrinos (LVD); nuclear 
astrophysics (LUNA). The laboratory is operated as an international science facility and 
hosts also non-INFN physics experiments, whose scientific value is assessed 
by an international advisory Scientific Committee. The laboratory is also supporting 
small-scale measurements on geology, biology and environmental issues. Many former experiments  
located at Gran Sasso - and now decommissioned - were at the leading frontier at that time. 
Among them, Gallex/GNO (solar neutrinos) and MACRO (atmospheric neutrinos and cosmic 
rays). \\
The laboratory hosts an underground low-background facility, STELLA (SubTErranean Low Level 
Assay), consisting of about 10 HPGe $\gamma$ spectrometers. Applications of the facility 
include material screening for experiments (radiopurity assay), background characterization, 
physics measurements (e.g. search for rare decays) and environmental measurements ($^{222}$Rn 
in ground water, radio-dating). The facility occupies a surface of about 32 m$^{2}$ in a 
connection tunnel and it hosts the most sensitive HPGe $\gamma$-spectrometer in the world, 
GEMPI2. The total background rate of the GEMPI2 
detector is 44~counts/(kg$\cdot$day) (integral between 40 and 2700~keV); 
the background rate at the $^{40}$K 1460-keV peak is 
0.13~counts/(kg$\cdot$day)~\cite{lau10,heu06}.
\subsubsection{Laboratorio Subterr\`aneo de Canfranc (LSC), Spain}
The first underground facility under the Mount Tobazo in the Spanish Pyrenees, near the French border,  
was created at the beginning of 80s, close to a dismissed railway tunnel. The laboratory was composed by three 
separated small halls along the railway tunnel, having a total surface of about 140 m$^{2}$ and 
maximum rock overburden ranging from 780 m w.e. to 2400 m w.e. 
The laboratory was operated by the University of Saragossa and fully characterized in terms of background. Several 
experiments were performed in the old Canfranc laboratory, notably IGEX ($0 \nu 2 \beta$ decay), ROSEBUD and 
ANAIS (dark matter searches). \\
 Taking profit from the 
excavation of a new parallel road tunnel, a new laboratory has been recently built. Excavation started 
in 2005 and the underground laboratory has been completed after the civil engineering on July 2010. The 
new Canfranc laboratory is managed by a Consortium including the Spanish Ministry for Education and Science, 
the Government of Aragon and the University of Saragossa. The new underground facility contains two main 
halls (labeled Hall A and Hall B) having surface of 600 m$^{2}$ ($40 \times 15 \times 12$~m$^{3}$) and 
150 m$^{2}$ 
($15 \times 10 \times 8$~m$^{3}$), respectively, and ancillary tunnels and services (e.g. clean room). 
Surface infrastructures, for a total area of about 1500 m$^2$ are also being built.  
The access is horizontal via one of the available road tunnels. The maximum rock overburden in the new 
laboratory is about 2400 m w.e. allowing a residual muon flux between  
2 and $4 \cdot 10^{-3}$~$\mu/($m$^2 \cdot$s)~\cite{bet07}, depending on the location. The new site still has to be 
characterized in terms of neutron flux, $\gamma$-ray flux and $^{222}$Rn activity\footnote{The fast neutron 
flux (above 
ca 1~MeV) measured in the old Canfranc laboratory is $(3.82 \pm 0.44) \cdot 10^{-2}$~n/(m$^2 \cdot$s)~\cite{car04}.}. The 
total ventilation power is about 11~000 m$^3$/h, corresponding to 1.5 volumes per hour. 
The underground laboratory is presently empty, but six experiments have been approved by the International 
Scientific Committee, on  $0 \nu 2 \beta$ decay (BiPo and NEXT), dark matter searches (ANAIS and ROSEBUD), 
low-background assays for liquid scintillators (SuperK-Gd) and geo-dynamics (GEODYN).  
\subsubsection{Boulby Palmer Laboratory (BUL), United Kingdom}
The Boulby Palmer Laboratory was established in 1998 and it is located under an active potash mine in 
the North-East of England operated by Cleveland Potash Ltd. The underground scientific infrastructure 
has a total area of about 1500 m$^{2}$, sub-divided in several tunnels, and is located at about 1100~m 
depth under a flat surface. Surface facilities (about 200 m$^2$) are also available, hosting ancillary 
services, as computing, chemical labs, mechanical workshop, electronic and offices. 
The access is vertical through a shaft operated by the mine company. The 
minimum rock overburden is 2850 m w.e. (average depth is 3450 m w.e.), giving 
a residual muon flux of $4.1 \cdot 10^{-4}$~$\mu/($m$^2 \cdot$s)~\cite{ara09}. The fast neutron flux above 
0.5~MeV has been 
measured to be $(1.72 \pm 0.72) \cdot 10^{-2}$~n/(m$^2 \cdot$s)~\cite{tzi07}. The particular composition of the 
rock, which is mainly salt, turns out in a very low $\gamma$-ray and $^{222}$Rn background. In particular, 
the abundance of primordial radionuclides in the laboratory salt is 67 ppb for $^{238}$U, 
125~ppb for $^{232}$Th and 1130~ppm for Potassium. The average $^{222}$Rn content in the underground site is 
less than 3~Bq/m$^3$. \\
The scientific activity of the Boulby laboratory is mainly focused on dark matter searches. The laboratory 
hosted the ZEPLIN-II experiment (completed in 2008) and is presently hosting its upgraded version, 
ZEPLIN-III, as well as the DRIFT-II R\&D. The future planning is still focused on dark matter searches but 
it pursues the further development of very low-background facilities for material screening and is open 
to future small and large projects, as LAGUNA. \\
The underground site hosts a ultra-low background HPGe spectrometer (about 2~kg weight), which is used 
for material activity measurements. Radon emanation measurements are routinely performed in the 
laboratory using low-background commercial detectors (Durridge Rad7), with sensitivity better than 
0.02~Bq/sample.
\subsubsection{Center for underground physics in  Pyh\"asalmi (CUPP), Finland}
The center is hosted since 2001 in a working mine close to Pyh\"asalmi, in central Finland.  
Several cavities, that were excavated between 1962 and  
2001 and have been dismissed by the mine, are available at different depths between 75 and 980 m.  
The total usable underground area is about 1000 m$^2$. New scientific facilities could be excavated 
at the depth of 1400 m (about 4 km w.e. minimum overburden), where the mining activities are currently 
taking place. Access is both via a vertical shaft and via a long inclined tunnel, which can be also 
accessed by trucks. The EMMA experiment on cosmic ray muons is currently hosted underground 
at the 75 m level. Small surface facilities are available, including offices and a guesthouse. The site 
is open for small and medium-scale experiments which may fit in the existing cavities. 
\section{Laboratories in Eastern Europe} \label{sec:otherlab}
In the following subsections, the four underground sites located in Eastern Europe 
will be described in more detail. As in the previous section, facilities are reported in 
chronological order since their establishment. Unless specifically quoted, fluxes and 
other information are taken from Ref.~\cite{bet07} and references therein.
\subsubsection{Baksan Neutrino Observatory (BNO), Russia}
The Baksan Neutrino Observatory, located under the Mount Andyrchi in the Russian Caucasus, is 
the oldest underground facility in the world which was explicitly built for scientific 
purposes. The excavation started in 1966 and experimental activities started in the 70s. A new 
village, called `Neutrino', was built as a part of the original project to host personnel and 
services (offices, water supply, heating). The laboratory is operated by the Institute for 
Nuclear Research (INR) of the 
Russian Academy of Sciences.  The underground facilities are located along two 
parallel horizontal tunnels, each 4 km long, excavated into the side of Mt. Andyrchi. The access is horizontal 
through the entrance tunnel, which is equipped with a rail system for the transportation of 
personnel and equipment. Several experimental halls are available, which are located at different 
distance from the entrance and have a different rock overburden. The BUST Hall 
($24 \times 24 \times 16$~m$^{3}$) has the least rock overburden (about 850 m w.e.) and hosts the 
BUST neutrino telescope, ready since 1978 to detect neutrinos from a galactic supernova explosion. The other main 
hall, having dimension $60 \times 10 \times 12$~m$^{3}$ and much larger overburden (4800 m w.e. 
vertical depth) hosts the SAGE solar neutrino experiment, which is operational since 1992. The SAGE 
site is lined with 60~cm of low-background concrete, to further reduce the neutron and $\gamma$-ray fluxes 
coming from the rock. The excavation of a larger and deeper hall, which started in 1990, was abandoned 
at the time of the collapse of the Soviet Union and its status is presently undefined. A few smaller 
halls with intermediate rock overburden are used for R\&D on dark matter and $0 \nu 2 \beta$ decay 
and host a screening facility with HPGe detectors. The laboratory is actually operated as an observatory, in the sense 
that its program foresees very long duration measurements. \\
The muon flux in the SAGE hall is $3.03 \cdot 10^{-5}$~$\mu/($m$^2 \cdot$s)~\cite{gav91}. The total neutron 
flux between 1 and 11 MeV is less than $2.3 \cdot 10^{-3}$~n/(m$^2 \cdot$s)~\cite{abd01}. 
The total $^{222}$Rn contamination in the laboratory is 
about 40 Bq/m$^{3}$, when the ventilation (7 volumes per hour) is on. 
\subsubsection{Solotvina Underground Laboratory (SUL), Ukraine}
The Solotvina underground laboratory is located in an active salt mine, at 430~m depth. The 
Laboratory was established in 1984 and it is operated by the Institute for Nuclear Research of 
the Ukrainian Academy of Sciences. The total underground 
area is about 1000~m$^2$, composed by a main hall ($25 \times 18 \times 8$~m$^{3}$) and four smaller 
halls, each $6 \times 6 \times 3$~m$^{3}$. The access to the laboratory is vertical through the 
mine shaft. A small surface facility is available with offices and living rooms. The rock profile is 
practically flat and the minimum overburden is about 1000 m w.e., giving a muon flux in the experimental hall 
of $1.7 \cdot 10^{-2}$~$\mu/($m$^2 \cdot$s)~\cite{zde88,dan95}. The thermal neutron flux is 
$< 2.7 \cdot 10^{-2}$~n/(m$^2 \cdot$s)~\cite{zde88}.  
The particular rock composition (salt) yields a lower $\gamma$-ray 
background with respect to laboratories excavated in ordinary rock, as discussed for the Boulby Laboratory, 
because of the lower concentration of primordial radionuclides of the $^{232}$Th and $^{238}$U series. The 
radon concentration in the laboratory air is 33~Bq/m$^3$. \\
The laboratory hosts a number of small- and medium-scale R\&D projects focused on $0 \nu 2 \beta$ decay 
(especially for SUPER-NEMO) and on the development of radiopure scintillating crystals 
for $0 \nu 2 \beta$ decay and dark matter experiments. Dedicated experiments on the $\beta\beta$ decay 
of $^{116}$Cd are also performed using pure $^{116}$CdWO$_4$ crystals. 
\subsubsection{Unirea mine, Romania}
Starting from 2006, scientific activities are ongoing underground in the Unirea salt mine, 
close to the Slanic town (Romania), in the sub-Carpatian hills. The mine has a hive-like structure 
composed by several galleries, each 32-36~m wide and 54-58~m high. The total floor area amounts 
to more than 70~000~m$^2$. A Low Background Radiation Laboratory 
with HPGe detectors is sitting at a depth of 208~m beneath the surface, corresponding to about 580 
m w.e.~\cite{mar08}. The access to the mine is vertical, through two elevators connected to the surface, 
each with 45-ton capacity. While the site is too shallow for most astroparticle physics experiments 
(dark matter search, $0 \nu 2 \beta$ decay) because of the $\mu$-induced background, it is an ideal 
place for a very low-background $\gamma$-spectrometry facility. In fact, the salt in the mine is 
extremely poor in primordial radionuclides of the $^{232}$Th and $^{238}$U series. The total dose measured 
in the underground area is $1.17 \pm 0.14$~nGy/h, which is about 80 times smaller than the value on 
the surface. No dedicated measurements have been performed up to now to evaluate the muon and neutron 
fluxes. The Unirea site is a possible candidate to host the LAGUNA initiative. A few HPGe detectors 
are presently operational underground.
%
\subsubsection{Polkowice-Sieroszowice mine, Poland}
The active Polkowice-Sieroszowice salt mine, located near Wroc{\l}aw, in the South-West of Poland, 
and operated by the KGHM holdings, is a candidate site for  
a new scientific underground laboratory in the near future (SUNLAB). The total excavated area covers 
many tens of km$^2$ and several large chambers in salt ($85 \times 15 \times 20$~m$^{3}$) are 
presently available at the depth of 950~m, corresponding to about 2200 m w.e. The access to the 
mine is vertical, with many shafts available. Measurements performed in the underground 
area with portable $\gamma$ spectrometers indicate very low levels of Uranium and Thorium, as 
expected for a salt mine, and a total dose of $1.9 \pm 0.14$~nGy/h~\cite{laguna}. The Radon content in 
the air is between 10 and 38 Bq/m$^3$, mainly due to the pumping of external air through the mine 
ventilation system. Muon and neutron backgrounds have not been measured yet, but dedicated activities for 
the background characterization of the site started in 2010. The Polkowice-Sieroszowice mine is a 
possible candidate to host the LAGUNA initiative.
\section{Conclusions}
Many underground sites are presently available for science in Europe with depth larger than 1 km w.e. 
The four main deep-underground laboratories in Western Europe (LNGS, LSM, Boulby and LSC) are very well 
established and all of them have a rich experimental program encompassing dark matter searches, neutrinoless 
double beta decay, neutrino physics and astroparticle physics. The four laboratories are the backbone of 
the EU deep underground scientific infrastructures. A closer coordination among them 
started in the framework of the ILIAS program, funded by the EU between 2004 and 2009. A memorandum of 
understanding (EULab) was signed after the end of ILIAS to improve the long-term laboratory 
coordination. \\
The Baksan underground laboratory, in Russia, is still operational, in spite of difficulties, 
mainly providing an observatory service to the scientific community. \\
Other ``emerging'' sites are becoming available, especially in Eastern Europe. While at the moment they lack 
of the necessary infrastructures to host full-scale physics experiments, their 
background is being  characterized and they are potential hosts for future physics programs.

\begin{theacknowledgments}
First of all, I would like to thank the Organizers of the LRT2010 workshop, and specifically 
R.~Ford and and N.~Smith, for their kind invitation. To visit the SNOLab facilities underground 
was indeed a new and unique experience.\\ 
While preparing the talk, I (re)used material and 
received updated information about the European laboratories from many Colleagues, whom I 
wish to thank here: A.~Bettini, E.~Coccia, M.~Laubenstein, L.~Miramonti, S.~Paling and 
F.~Piquemal. I am indebted to V.~Kudryavtsev for many enlightening discussions on muon 
flux in underground laboratories and on the calculation of the averaged depth.
\end{theacknowledgments}


\end{document}